\documentclass[aps,apl,twocolumn,groupaddress]{revtex4}
\usepackage{epsfig}
\DeclareMathAlphabet{\mathitb}{OT1}{cmr}{bx}{sl}

\begin{document}

\renewcommand{\thefootnote}{\fnsymbol{footnote}}

\title{Electron Number Dependence of Spin Triplet-Singlet Relaxation Time}

\author{Gang Cao}
\author{Ming Xiao}
\email{maaxiao@ustc.edu.cn}
\author{HaiOu Li}
\author{Tao Tu}
\author{GuoPing Guo}
\email{gpguo@ustc.edu.cn}
\affiliation{Key Laboratory of Quantum Information, Chinese Academy of Sciences, University of Science and Technology of China, Hefei 230026, People's Republic of China}
\date{\today}

\begin{abstract}
In a GaAs single quantum dot, the relaxation time $T_{1}$ between spin triplet and singlet states has been measured for the last few even number of electrons. The singlet-triplet energy separation $E_{ST}$ is tuned as a control parameter for the comparison of $T_{1}$ between different electron numbers. $T_{1}$ shows a steady decrease from 2-electrons, 4-electrons, to 6-electrons, and we found this implies an enhancement of the spin-orbital coupling strength in a multi-electron quantum dot.
\end{abstract}

\maketitle

The spin singlet-triplet states of an electron pair in a quantum dot have been demonstrated as potential solid-state qubits \cite{Spin-Qubits-Kouwenhoven, Spin-Qubits-Petta, Spin-Qubit-Tarucha}.  Principally, any even number of electrons would form the singlet-triplet configuration. Experimentally, spin blockade effect for a variety of even number of electrons have been observed \cite{Spin-Blockade-Multi-Electrons} and the coherent manipulation of multi-electron singlet-triplet-based qubits has been recently studied \cite{Spin-Qubit-Electrons-APS}. A question that naturally arises is whether the multi-electron interaction interferes with the singlet-triplet coherence, such as enhancing relaxing or dephasing.

Here we study the singlet-triplet relaxation time $T_{1}$ in a single quantum dot for different even number of electrons. Since $T_{1}$ strongly depends on the singlet-triplet energy separation $E_{ST}$, we control $E_{ST}$ by tuning the quantum dot shape with confinement gates. For a given electron number, $T_{1}$ is measured with the pump-and probe technique \cite{T1-Pump-and-Probe} in a tunable range of $E_{ST}$. $T_{1}$ shows large decrease (roughly 3 times) when the electron number increases from $2$ to $6$. An increase in the spin-orbit coupling strength with larger electron number is found to explain the observed decrease in $T_{1}$.

\begin{figure}[t]
\begin{center}
\epsfig{file=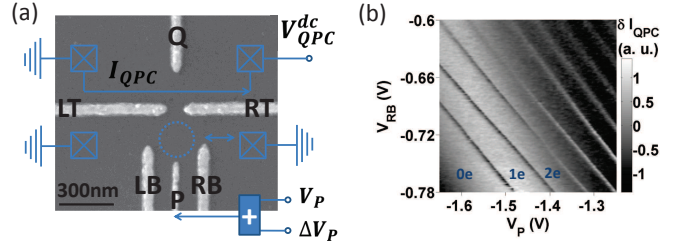, width=1\linewidth, angle=0}
\end{center}
\vspace{-7.5mm}
\caption{(a) A SEM image showing the geometry of our sample. The dotted circle is the location of the quantum dot. Gate $P$ is used to control the QD electron energy with respect to the Fermi level of the electron reservoir. Sometimes voltage pulse $\Delta V_{P}$ will be applied on gate $P$ to dynamically explore the quantum dot energy spectrum. (b) Gray-scale plot of the QPC differential current as functions of voltages $V_{P}$ and $V_{RB}$. Voltages on other gates are: $V_{LB}$ = -1.40 V, $V_{LT}=V_{RT}$ = -1.50 V, $V_{Q}$ = -0.90 V, and $V_{QPC}^{dc}$ = 0.3 mV.}
\label{Figure1}
\end{figure}

Fig. \ref{Figure1}(a) shows a scanning electron microscopy (SEM) image of the gate-defined single GaAs quantum dot. The left barrier of the dot is closed and the electrons only tunnel through the right barrier. The dc current through the quantum point contact (QPC) is recorded to count the charge number on the dot. A gap between the QPC and the dot is created to maximize the charge counting sensitivity. In this experiment the gap is closed tightly and the QPC dc bias voltage $V_{QPC}^{dc}$ is small to minimize the back-action effect \cite{Backaction-Kouwenhoven, Ming-Backaction-RTS}. Fig. \ref{Figure1}(b) shows the charge stability diagram measured by the QPC while gate $P$ and $RB$ are biased at dc voltages. We will measure $T_{1}$ of the spin singlet-triplet states for $2e$, $4e$, and $6e$, respectively.

\begin{figure}[t]
\begin{center}
\epsfig{file=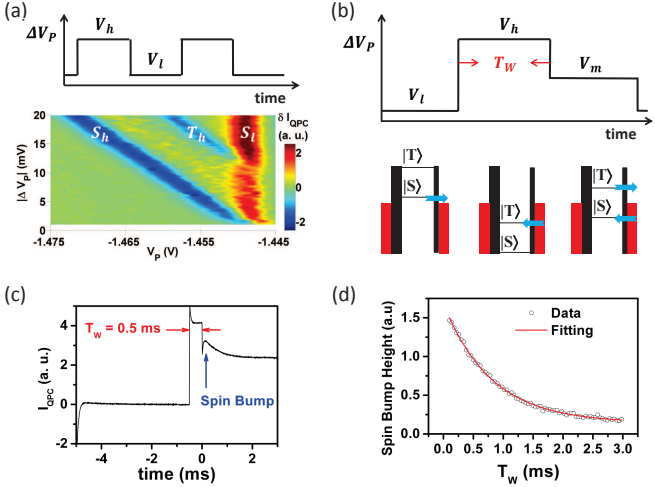, width=1\linewidth, angle=0}
\end{center}
\vspace{-7.5mm}
\caption{(a) A sequence of square-waves is applied on gate $P$. The pulse frequency is typically 600 Hz. The gray-scale plot shows the numerically differentiated QPC current measured by a lock-in amplifier with time constant 300 ms. This graph is taken around the 1e $\leftrightarrow$ 2e transition region. $V_{RB}=$ -0.76 V and all other gate voltages are the same as in Fig. \ref{Figure1}(b). (b) The mechanism of the pump-and-probe measurement for the spin relaxation time when a sequence of three-step pulses is applied on gate $P$. (c) The gate-averaged QPC current over a sequence of 4000 pump-and-probe pulses. It begins with the low-level pulse, followed by a high-level pulse ($T_{W}=$ 0.5 ms in this example). Finally from 0 ms to 3 ms, the spin-bump is read in the medium-level pulse step. (d) The spin-bump height as a function of $T_{W}$. Open dots are the experimental data. Solid curve is the fitting with an exponential decay.}
\label{Figure2}
\end{figure}

As shown in Fig. \ref{Figure2}(a), a sequence of square-wave voltage pulses is applied on the plunger gate $P$ to explore the energy spectroscopy of the quantum dot by pumping the electrons to excited states \cite{Spectroscopy-Two-Pulse}. The gray-scale plot shows the  QPC response averaged over many duty cycles with a lock-in amplifier, in the 1e $\leftrightarrow$ 2e transition region. During each duty cycle, the low-level pulse $V_{l}$ and high-level pulse $V_{h}$ bring the QD electrons into the spin ground state $|S\rangle$ twice, and correspondingly produce two charge transition lines, denoted as $S_{h}$ and $S_{l}$. When the pulse amplitude is large enough, the high-level pulse $V_{h}$ pumps the electrons into the spin excited state $|T\rangle$ as well. In fact, we see an additional line denoted as $T_{h}$ between $S_{h}$ and $S_{l}$ when $|\Delta V_{P}| \equiv |V_{h}-V_{l}| \geq$ 12.3 mV. Using the energy-voltage conversion factor 0.07 meV/mV, we determined $E_{ST}$ as 0.86 meV in this example. 

In order to detect the relaxation process from $|T\rangle$ to $|S\rangle$, we applied a sequence of three-step pulses \cite{T1-Pump-and-Probe, T1-Three-Pulse}, as illustrated in Fig. \ref{Figure2}(b). At the low voltage level $V_{l}$, the energy of both $|T\rangle$ and $|S\rangle$ lie above the Fermi level $E_{F}$ of the electron reservoir. This resets the quantum dot by emptying out the $|S\rangle$ state. Then at the high voltage level $V_{h}$,  the energy of both $|T\rangle$ and $|S\rangle$ drop below $E_{F}$. Therefore the electrons are pumped into the spin excited state $|T\rangle$ with a certain probability. $V_{h}$ sustains for a waiting time $T_{W}$, during which period $|T\rangle$ relaxes to $|S\rangle$. Finally at the medium voltage level $V_{m}$, $E_{F}$ is brought between the energy of $|T\rangle$ and $|S\rangle$. If the electrons have already relaxed into $|S\rangle$, no electron jumping will occur in such an energy configuration. On the other hand, if the relaxation has not completed and the electrons still have a probability of occupying state $|T\rangle$, one electron will jump out of the dot and another electron will jump in to fill state $|S\rangle$. As a consequence, the QPC current shows a "up" and "down" switching. So the occurrence probability of the QPC current switching at the $V_{m}$ level, as a function of $T_{W}$, can tell us the speed of the $|T\rangle - |S\rangle$ relaxation process.

Repeating the above process many times, the "up" and "down" switchings of the QPC current at the $V_{m}$ level average out as Gaussian distributions, called a "spin bump". The dependence of the spin bump height on $T_{W}$ contains the information of $T_{1}$. Fig. \ref{Figure2}(c) shows a typical trace of the gate-averaged QPC current recorded by a high-bandwidth oscilloscope. The spin bump occurs when the pulse time is from 0 ms to 3 ms. We increase the waiting time $T_{W}$ and see that the spin bump height rapidly decays. Fig. \ref{Figure2}(d) shows the extracted dependence of the spin bump height on $T_{W}$, which can be described by an exponential decay with a rate $1/T_{1}$. The fitting in Fig. \ref{Figure2}(d) gives $T_{1}=0.88\pm0.01$ ms.

It is well known that $T_{1}$ strongly depends on the energy splitting $E_{ST}$, because of the energy dependence of the phonon emission rate and the electron-phonon interaction \cite{T1-Phonons-DLoss, T1-EST-B-Vt-Delft, T1-ST-Mso, T1-Quantitative-Model}. $E_{ST}$ keeps increasing when we sweep down $V_{P}$ to squeeze the electron number from 6 to 2, because the dot size shrinks and the exchange interaction energy increases. During this process, a longer $T_{1}$ is observed for a smaller electron number. But in this situation we can not tell if the change in $T_{1}$ comes from the difference in $E_{ST}$ or in the electron number. A meaningful comparison can only be made at the same value of $E_{ST}$ while the electron number is varied.

\begin{figure}[t]
\begin{center}
\epsfig{file=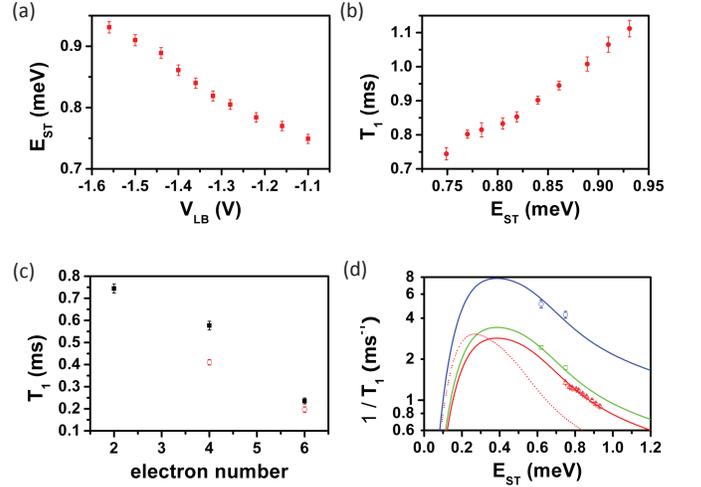, width=1\linewidth, angle=0}
\end{center}
\vspace{-7.5mm}
\caption{(a) Tuning $E_{ST}$ with gate voltage $V_{LB}$ for $2e$. (b) For $2e$, $T_{1}$ as a function of $E_{ST}$. (c) $T_{1}$ for different electron numbers. $E_{ST}$ is fixed at $0.75\pm0.01$ meV for the black closed squares, and $0.62\pm0.01$ meV for the red open circles. (d) $1/T_{1}$ as a function of $E_{ST}$. The red triangles, green squares, and blue circles are for $2e$, $4e$, and $6e$, respectively. The solid curves are the theoretic fittings. The red dotted curve is the same fitting with the red solid curve, except that the dot radius is enlarged from 16 nm to 20nm.}
\label{Figure3}
\end{figure}

The control experiment on $E_{ST}$ can be done with tuning some of the confinement gates, mainly gate LB. When increasing $V_{LB}$ (and slightly compensating $V_{RB}$), we found that $E_{ST}$ continuously decreases, as shown in Fig. \ref{Figure3}(a) for the case of $2e$. This control on $E_{ST}$ is most likely fulfilled by changing the quantum dot shape \cite{T1-EST-B-Vt-Delft, Zeeman-Vshape-Kastner}. A more positive voltage on gate LB pulls the electron wave functions on one side of the dot and makes the dot less circular. This gives rise to smaller $E_{ST}$. In this experiment, we can tune $E_{ST}$ from 0.93 to 0.75 meV for 2e by increasing $V_{LB}$ over 460 mV.

When $E_{ST}$ goes up, $T_{1}$ monotonically increases, as can be seen in Fig. \ref{Figure3}(b) for $2e$. For $4e$ and $6e$,  we also observed a monotonically increase of $T_{1}$ with $E_{ST}$. However it becomes more and more difficult to collect enough data points for larger electron numbers. When $V_{LB}$ is the same, $E_{ST}$ for $4e$ and $6e$ is usually about one half of $E_{ST}$ for $2e$. It is therefore not easy for $E_{ST}$ to reach the same magnitude for different electron numbers. An exhaust tuning of one confinement gate at relatively large dot size turns out to result in double or multiple dots. Nonetheless, we managed to bring up $E_{ST}$ for $4e$ and $6e$ to reach the lower limit of $E_{ST}$ for $2e$. In Fig. \ref{Figure3}(c) we present $T_{1}$ for $2e$, $4e$, and $6e$ for two fixed values of $E_{ST}$, 0.75 meV and 0.62 meV, respectively. We see that at each fixed value of $E_{ST}$, $T_{1}$ substantially decreases with increasing electron number. For example, when $E_{ST}$ = 0.75 meV, $T_{1}$ drops from 0.75 ms for $2e$ to only 0.23 ms for $6e$, which is a change of more than 3 times. 

To quantitatively evaluate the change in $T_{1}$, we use the simplified model \cite{T1-EST-B-Vt-Delft, T1-Quantitative-Model}:
\begin{center}
$1/T_{1} = \frac{M_{SO}^{2}}{32\pi \rho \hbar^{3}} \Big( \frac{\Xi^{2} a^{4}}{\lambdabar_{l}^{5} c_{l}^{4}} {\displaystyle \int_{0}^{\pi/2}} d\theta sin^{5}\theta e^{-(a^2 sin^2\theta) / (2 \lambdabar_{l}^{2})}$
$+ \sum\limits_{j=l,t} \frac{e^{2} \beta^{2} a^{4}}{\lambdabar_{j}^{3} c_{j}^{4}} {\displaystyle \int_{0}^{\pi/2}} d\theta |A_{j}(\theta)|^{2} sin^{5}\theta e^{-(a^2 sin^2\theta) / (2 \lambdabar_{j}^{2})} \Big) $
\end{center}

Here the first integral comes from the deformation potential electron-phonon coupling and is nonzero only for the longitudinal phonons. The second integral is from the piezoelectric coupling and contains contributions from both the longitudinal and transverse phonons. The piezoelectric constants are as following: $A_{l}(\theta)=3 \sqrt{2} sin^{2}(\theta) cos\theta /4$, $A_{t_{1}}(\theta)=\sqrt{2} sin(2\theta) /4$, and $A_{t_{2}}(\theta)=\sqrt{2} (3cos^{2}(\theta)-1) sin\theta /4$. The coupling strength for the deformation potential and piezoelectric interactions is $\Xi$ = 6.7 eV and $e\beta = 1.4\times10^{9}$ eV/m, respectively \cite{T1-Phonons-DLoss, T1-EST-B-Vt-Delft}. The mass density $\rho$ is 5300 kg/m$^{3}$, and the sound speed for the longitudinal and transverse phonons is $c_{l}$ = 4730 m/s and $c_{l}$ = 3350 m/s, respectively \cite{T1-EST-B-Vt-Delft}. 

The dot radius $a$ is estimated as about 16 nm using transport experiments. The characteristic energy, where the phonon wavelength $\lambdabar_{l,t} = \hbar c_{l,t} / E_{ST}$ matches the dot size and the relaxation rate $1/T_{1}$ reaches a maximum, is therefore $E_{ST}^{*} = \hbar c_{l,t} / a =$ 0.19 or 0.14 meV. This is much smaller than the energy range in our experiment, and we see from Fig. \ref{Figure3}(d) that our data points lie to the right side of the peak points in the simulated curves.

The spin-orbit strength $M_{SO}$ is an independent factor in our simulation. A value 0.33 $\mu$eV gives a good match to our data for $2e$, shown as the red curve and red symbols in Fig. \ref{Figure3}(d). This is consistent with the reported 0.37 $\mu$eV in the literature \cite{T1-EST-B-Vt-Delft}. $M_{SO}$ also shows a dependence on the electron number. The green (blue) curve in Fig. \ref{Figure3}(d) tells us that $M_{SO}$ for 4e (6e) is 0.36 (0.55) $\mu$eV, which is about 1.1 (1.7) times of $M_{SO}$ for 2e. These changes result in measurable increase in the relaxation rate $1/T_{1}$ due to its quadratic dependence on $M_{SO}$. The dot radius $a$ may also increase with larger electron number. In Fig. \ref{Figure3}(d) the red dotted curve is the predicted $1/T_{1}$ for $2e$ with larger $a$ (20 nm) and same $M_{SO}$ (0.33 $\mu$eV) as the red solid curve. The peak position shifts to the left and the magnitude decreases on the right side of the peak. So the enlargement of dot size seems unable to explain the observed increase of $1/T_{1}$. 

It has been pointed out that the Coulomb interaction between electrons has large influence on a number of factors in the singlet-triplet relaxation process, and most likely induces stronger spin to charge coupling \cite{T1-ST-Mso}. So the multi-electron Coulomb interaction may be responsible for the enhancement of the spin-orbit coupling strength in a multi-electron quantum dot. The inclusion of more orbital states could possibly strengthen the spin-orbital coupling too.

In conclusion, we studied the relaxation time of spin singlet-triplet states for the last few even electron numbers. With increasing electron number, even only from 2 to 6, $T_{1}$ was found to substantially decrease, possibly due to the enhanced spin-orbital coupling strength. Therefore the exact electron number may be a matter when we choose a singlet-triplet as the basis of a qubit.

This work was supported by the NFRP 2011CBA00200 and 2011CB921200, and NNSF 10934006, 11074243, 11174267, 91121014, and 60921091.

\end{document}